\documentclass[11pt,draftclsnofoot,onecolumn]{IEEEtran}
\IEEEoverridecommandlockouts
\usepackage[utf8]{inputenc} 
\usepackage[T1]{fontenc}
\usepackage{url}              
\usepackage{cite}             
\usepackage[cmex10]{amsmath}
\usepackage{amssymb, amsthm} 
\usepackage{graphicx}         
\usepackage{subfigure}        
\usepackage{mathtools}        
\usepackage{float}            
\usepackage{algorithm}        
\usepackage{algorithmic}      
\usepackage[colorlinks=true, linkcolor=blue, citecolor=green]{hyperref} 
\usepackage{multirow}         
\usepackage{tikz}             
\usepackage{booktabs}         
\usepackage{stackrel}
\usepackage[top=0.75in,bottom=1in,left=0.625in,right=0.625in]{geometry}
\usetikzlibrary{%
	arrows,%
	shapes.misc,
	shapes.arrows,%
	chains,%
	matrix,%
	positioning,
	scopes,%
	decorations.pathmorphing,
	shapes.geometric,
	shadows,
	decorations.pathreplacing,
	patterns,
	shapes,
	decorations.markings
}

\newcommand{\midarrow}{\tikz \draw[-triangle 45] (0,0) -- +(0,0.1);}

\newtheorem{defi}{Definition}[section]
\newtheorem{thm}{Theorem}[section]
\newtheorem{lem}{Lemma}[section]
\newtheorem{corol}{Corollary}[section]

\newcommand{\Ac}{\mathcal{A}}

\newcommand{\Pc}{\mathcal{P}}
\newcommand{\Xc}{\mathcal{X}}

\newcommand{\Mc}{\mathcal{M}}

\newcommand{\Nc}{\mathcal{N}}

\newcommand{\EWR}[2]{\mathbb{E}_{#1} \left [ #2 \right ]}

\newcommand{\pr}[1]{\Pr \left\{ #1 \right\}}

\newcommand{\lrpar}[1]{\left( #1 \right)}

\newcommand{\lrbrc}[1]{\left\{#1\right\}}

\begin{document}
	
	\title{Over-the-Air Federated Adaptive Data Analysis: Preserving Accuracy via Opportunistic Differential Privacy}
	
	\author
	{
		Authors\\
		Sharif University of Technology, Tehran, Iran
		Email: \{Authors\}@ee.sharif.edu}
	
	\author{Amir Hossein Hadavi, Mohammad M. Mojahedian and~Mohammad Reza Aref\\
		
		Information Systems and Security Lab (ISSL)\\
		
		Department of Electrical Engineering, Sharif University of Technology, Tehran, Iran\\
		
		Email: \{ah{\_}hadavi\}@ee.sharif.edu,  \{m.mojahedian\}@gmail.com,  \{aref\}@sharif.edu
		
		\thanks{}}
	
	\maketitle

\begin{abstract}
Adaptive data analysis (ADA) involves a dynamic interaction between an analyst and a dataset owner, where the analyst submits queries sequentially, adapting them based on previous answers. This process can become adversarial, as the analyst may attempt to overfit by targeting non-generalizable patterns in the data. To counteract this, the dataset owner introduces randomization techniques, such as adding noise to the responses. This noise not only helps prevent overfitting, but also enhances data privacy. However, it must be carefully calibrated to ensure that the statistical reliability of the responses is not compromised. In this paper, we extend the ADA problem to the context of distributed datasets. Specifically, we consider a scenario where a potentially adversarial analyst interacts with multiple distributed responders through adaptive queries. We assume the responses are subject to noise, introduced by the channel connecting the responders and the analyst. We demonstrate how this noise can be opportunistically leveraged through a federated mechanism to enhance the generalizability of ADA, thereby increasing the number of query-response interactions between the analyst and the responders. We illustrate that the careful tuning of the transmission amplitude based on the theoretically achievable bounds can significantly impact the number of accurately answerable queries.
\end{abstract}

\IEEEpeerreviewmaketitle

\section{Introduction}
In the era of big data, the necessity for systems that can dynamically and efficiently analyze vast and complex datasets has never been greater.One prominent approach in this domain is Adaptive Data Analysis (ADA). In ADA, the analyst aims to derive insights about an unknown distribution over a given domain by adaptively exchanging query-answer pairs with a responding mechanism that possesses a dataset of samples drawn from that distribution. However, theoretical considerations of ADA in contexts involving distributed datasets are relatively unexplored despite the growing prevalence of such datasets in modern computing environments.This paper proposes implementing ADA in distributed systems as an extension of Federated Learning (FL), which traditionally focuses on using gradient descent queries to train deep neural networks across decentralized locations. The following subsections provide a necessary examination of ADA and FL.

\subsection{Adaptive Data Analysis} \label{ADA}
In the basic model of ADA, there is one analyst ($\Ac$) and one answering mechanism ($\Mc$), communicating with each other through error-free channels with unlimited capacities. Hence, there is no concern about practical communication limitations in this basic form of the problem. It has assumed that the answering mechanism $\mathcal{M}$ has access to the dataset $S$ with $n$ samples generated from the domain $\mathcal{X}$ by the distribution $\mathcal{D}_{\mathcal{X}^n}$. Albeit, in the general case, $\mathcal{D}_{\mathcal{X}^n}$ is not necessarily a product distribution, in the sequel, we add this assumption as it is the case in most of the basic papers in the field. So, we assume that $\mathcal{D}_{\mathcal{X}^n} \equiv \mathcal{P}^n$, where $\mathcal{P}$ is a distribution on $\mathcal{X}$. The analyst $\mathcal{A}$ does not have direct access to $S$,  but can query $\mathcal{M}$ about it.Various families of queries are considered in the ADA papers, among which the most basic one is the statistical queries, studied in the pioneering work \cite{dwork2015preserving} and also in several subsequent works such as \cite{dwork2015generalization, rogers2020guaranteed}. In this paper, we also focus on this class of queries, which are formally defined below.
\begin{defi}
	A statistical query is uniquely specified by a function like $q : \Xc \rightarrow [0,1]$ and its answer is the linear average of the random samples, drawn from a particular distribution. More exact, It's answer with respect to the distribution $\Pc$ will be
	\begin{equation}
		q(\Pc)=\EWR{x\sim\Pc}{q(x)}.
	\end{equation}
\end{defi}
Despite their simplicity, statistical queries encompass a substantial range of practical applications. For instance, gradient descent updates can be viewed as a vector of statistical queries. Notably, most FL papers focus on gradient descent queries due to their pivotal role in the learning process of deep neural networks. Consequently, our work can be regarded as an abstraction of these studies.

The analyst $\mathcal{A}$ can adaptively make queries based on the previous question-answer pairs. More precisely, the $i$th query, $q_i$, is formulated considering the sequence $(q_1, a_1, \dots, q_{i-1}, a_{i-1})$, where $a_j$ is $\mathcal{M}$'s answer to the $j$th question, using the dataset $S$. The goal is to ensure that the empirical answers remain close to the true answers derived from the underlying distribution. The following definition provides the exact formulation of the accuracy criterion.
\begin{defi}
	We say that the adaptive data analysis procedure is $ (\alpha,\beta) $-accurate if:
	\begin{equation}\label{accdef}
		\pr{\max_i{ \left\vert  q_i(\mathcal{P}) - a_i \right\vert } \geq \alpha} \leq \beta,
	\end{equation}
	in which $q_i(\mathcal{P})$ is the expected value of the answer to the $i$'th query and the probability is calculated on all of the randomnesses of the model.
\end{defi}
In the basic model, because $\Mc$ does not know the action model of $\Ac$, it considers the worst-case scenario. Expressly, $\Mc$ assumes that $\Ac$ selects queries adversarially, aiming to overfit to the dataset $S$ and thereby violate the accuracy criterion defined by \eqref{accdef}. On the other hand, $\Mc$ aims to answer as many queries as possible while ensuring that \eqref{accdef} remains satisfied. To achieve this, $\Mc$ seeks an efficient answering mechanism. In the case of non-adaptive queries (i.e., queries selected independently of observations from the data), a natural unbiased approach for $\Mc$ would be to respond using the empirical answers computed from $S$. Nevertheless, there are well-known examples demonstrating that this approach can be inefficient in adaptive scenarios. For example, see \cite{dwork2015preserving}. A widely adopted strategy in the research community for addressing the adaptive case is to introduce randomness into empirical answers by adding stochastic noise. The idea of adding noise comes from the data privacy framework, particularly differential privacy, which serves as a privacy-enhancing mechanism \cite{dwork2014algorithmic}. In this approach, i.i.d. noise samples from a suitable distribution are generated and added to each empirically computed answer. The variance of the noise is calibrated to ensure that the accuracy criterion in \eqref{accdef} is satisfied. The noise distribution is typically chosen to be either Gaussian or Laplace. In this paper, we focus on the Gaussian distribution, as we aim to leverage Gaussian channel noise as the source of randomization. 

Since 2015, numerous studies have examined various aspects of ADA. In this paper, we apply one of the foundational results to our approach, rather than delving into the theoretical details. To our knowledge, no previous work has investigated the physical channel aspects of this problem. Moreover, this study is the first to explore the federated scenario within the context of ADA.

\subsection{Federated learning}
As previously noted, prior studies have not explored the physical and distributed aspects of ADA. However, the notion of distributed responders introduced in this work closely aligns with concepts in FL. To provide context, we cite selected works from the extensive FL literature that explore ideas analogous to those studied here.

In FL, a central node, called the Parameter Server (PS), interacts with multiple edge nodes, or clients. Together, these nodes work to train a shared neural network. Each client uses its local dataset to contribute to the training process. At the start of each training round, the PS broadcasts the updated weights to all nodes. Each node computes the gradient of the loss function using the current parameter values and its local dataset. Finally, the nodes send their local gradient estimations to the PS, which aggregates them to produce a more accurate global estimation.

The literature proposes various channel models and communication methods for the gradient transmission step. These approaches are generally divided into two categories: analog transmission and digital transmission. In this paper, we adopt the analog approach. With analog transmission, clients simultaneously send their estimates to the PS using uncoded transmission. This method is bandwidth-efficient because the channel is shared. Moreover, the additive property of the medium allows local estimations to naturally aggregate in the air. While this approach offers computational and storage efficiency, the aggregated signal may contain more noise and interference than digital transmission. However, these effects are not inherently detrimental. Some FL studies have utilized channel distortions to enhance client data privacy. Furthermore, \cite{yang2021revisiting} demonstrates that, although channel imperfections can slow the convergence rate, they may also improve the accuracy of the trained network. This suggests a nuanced relationship between data privacy and algorithm accuracy.

In the context of FL, the channel model is a critical consideration, particularly the availability of Channel State Information (CSI).
Several studies, such as \cite{amiri2020federated, mohamed2021privacy, sonee2021wireless, seif2020wireless}, assume perfect global CSI at the PS and perfect local CSI at the edge devices. Some Others, including \cite{xiaowen2021optimized, koda2020differentially, koda2021airmixml, liu2020privacy, zhang2021turning, OBDAgunduz}, focus solely on perfect local CSI at the edge devices. In contrast, \cite{shao2021federated} assumes that each edge device has imperfect knowledge of its channel gain to the PS, resulting in some mismatch between the estimated and actual channel gain. Another form of imperfect CSI is examined in \cite{OBDAgunduz, amiri2020blind}, where the PS is aware only of the sum of the fading coefficients, and the edge devices lack any CSI knowledge. Finally, \cite{yang2021revisiting} proposes a scheme that requires no CSI at all.

Another significant factor is the channel's fading model. The most basic model, adopted by works such as \cite{amiri2020machine, lee2021over}, uses the Gaussian MAC model, which does not account for the channel's attenuating behavior.

For a more practical approach, several works, including \cite{liu2020privacy,xiaowen2021optimized,koda2021airmixml}, consider fading effects, specifically the block-flat fading regime, where the channel condition remains constant during each communication round and may change afterward. 
In \cite{amiri2020federated}, fast fading is considered.  
Also \cite{koda2020differentially} incorporates both path loss and small-scale fading into the model, assuming that the EPs have perfect knowledge of both types of losses.

\subsection{Contributions and Paper Structure}
In this study, we explore, for the first time, the execution of ADA over noisy communication channels. As noted earlier, the ADA framework typically introduces independent randomnesses to answers to preserve statistical validity. Thus, instead of treating channel noise as a drawback, we propose leveraging it as a natural substitute for artificially added noise.

Furthermore, we extend this approach to a distributed scenario, where an analyst—potentially adversarial—interacts with multiple EPs via a sequence of queries and responses rather than working with a single answerer. The responses are transmitted through an Additive White Gaussian Noise (AWGN) channel. We propose a response mechanism that exploits the channel noise to enhance the generalizability of ADA and optimize the number of query-response interactions.

Section \ref{sec:sys_model} outlines the system model, while Section \ref{sec:P2P_basic_ADA} presents the core theoretical materials. Section \ref{sec:P2P_ADA} examines the ADA problem with a single dataset, considering the effects of the AWGN channel. Section \ref{sec:distributed_ADA} expands the discussion to ADA involving multiple distributed datasets. Finally, Section \ref{sec:conclusion} explores potential future directions and concludes the paper.

\section{System Model}
\label{sec:sys_model}
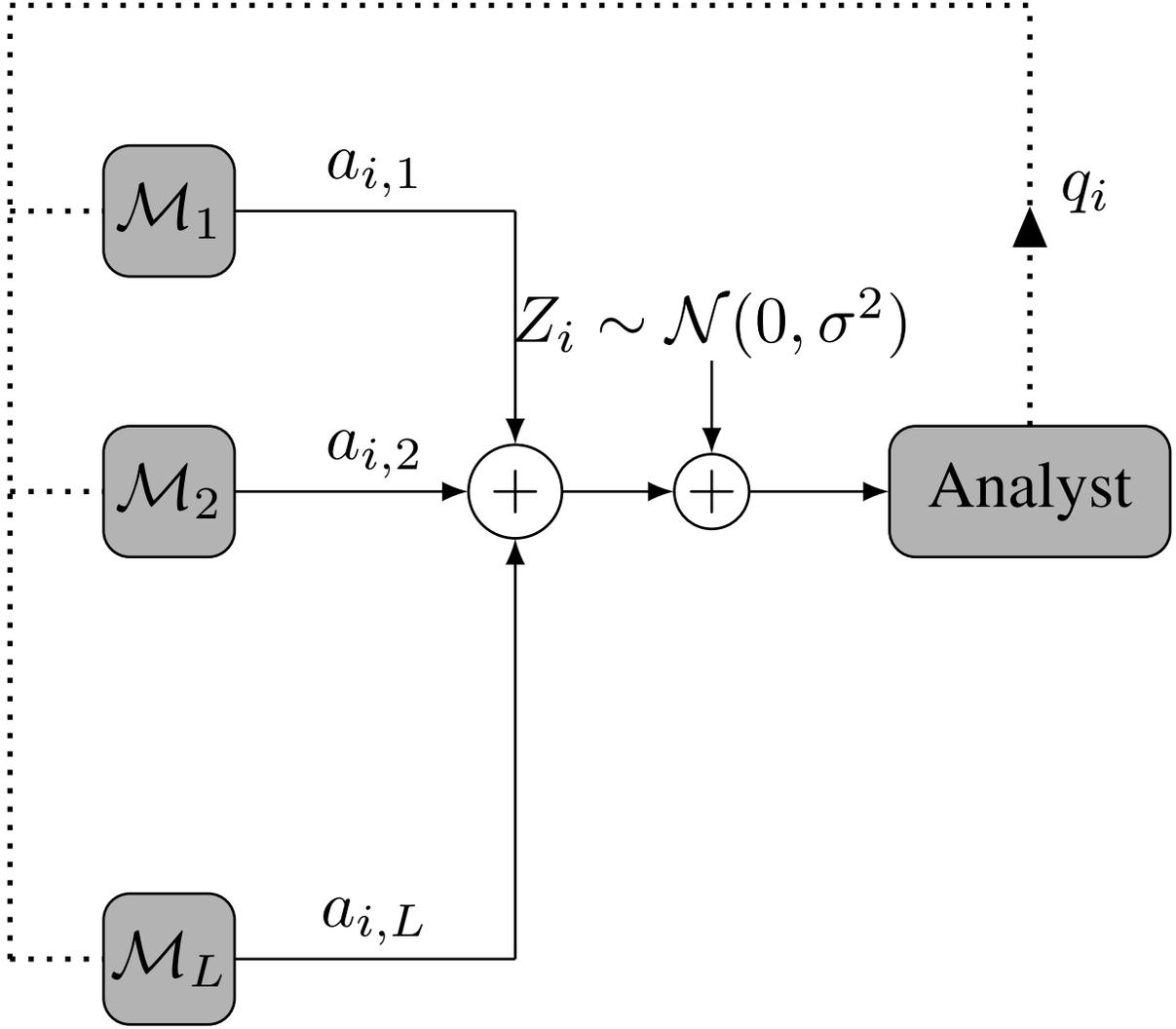
\begin{figure}[htbp]
	\centering
	\resizebox {.9\columnwidth} {!} {
		\begin{tikzpicture}[scale=1]
			\tikzstyle{every node}=[font=\small]
			
			\draw[dotted,thick] (-0.2,-1.5) -- (-0.2,3.6);
			\draw[dotted,thick] (-0.2,3.6) -- (5.25,3.6);
			
			\draw[dotted,thick] (-0.2,2.5) -- (0.3,2.5);
			\draw[dotted,thick] (-0.2,1) -- (0.3,1);
			\draw[dotted,thick] (-0.2,-1.5) -- (0.3,-1.5);
			
			\draw[dotted,thick] (5.25,1.35) --node {\midarrow} (5.25,3.6) node at (5.55,2.6) {$q_i$};

			\draw[rounded corners,fill=black!30!white] (0.3,2.15) rectangle (1,2.85) node [align=center,pos=.5] {$\mathcal{M}_1$};
			
			\draw (1,2.5) -- (2.5,2.5) node at (1.75,2.7) {$a_{i,1}$};
			\draw[-latex] (2.5,2.5) -- (2.5,1.25);
			
			\draw[rounded corners,fill=black!30!white] (0.3,0.65) rectangle (1,1.35) node [align=center,pos=.5] {$\mathcal{M}_2$};
			
			\draw[rounded corners,fill=black!30!white] (4.5,0.65) rectangle (6,1.35) node [align=center,pos=.5] {Analyst};
			\draw[-latex] (3.75,1) -- (4.5,1);

			\draw[-latex] (1,1) -- (2.25,1) node at (1.75,1.2) {$a_{i,2}$};

			\draw[rounded corners,fill=black!30!white] (0.3,-1.85) rectangle (1,-1.15) node [align=center,pos=.5] {$\mathcal{M}_L$};
			
			\draw (1,-1.5) -- (2.5,-1.5) node at (1.75,-1.3) {$a_{i,L}$};
			\draw[-latex] (2.5,-1.5) -- (2.5,0.75);

			\draw (2.5,1) circle (0.25) node at (2.5,1) {$+$};
			
			\draw (3.55,1) circle (0.2) node at (3.55,1) {$+$};
			
			\draw[-latex] (2.75,1) -- (3.35,1);
			\draw[-latex] (3.55,1.7) -- (3.55,1.2) node at (3.55,1.9) {$Z_i\sim\mathcal{N}(0,\sigma^2)$};
		\end{tikzpicture}
	}
	\caption{We have $L$ responders that respond to analyst queries through an AWGN channel.}
	\label{fig:ADA}
\end{figure}

As depicted in Figure \ref{fig:ADA}, a distributed situation of ADA is considered. In this problem, we assume that each of the $L$ Edge Points (EPs) $\mathcal{M}_1,\mathcal{M}_2,\dots,\mathcal{M}_L$  are connected to a Central Point (CP), referred to as the analyst, via a communication channel with limited capacity. Each EP, such as $\mathcal{M}_l$, holds a local dataset $S_l$ containing $n_l$ data samples and functions as an answering mechanism. The CP broadcasts its queries via error-free channels to all or some of them, and receives the answers from them through a wireless MAC. Note that this channel model is similar to those proposed in many recent FL studies, some of which were referenced in the previous section. This scenario can model a network of sensors communicating with a CP to infer environmental characteristics using data samples collected by the sensors.

In our proposed scheme, the answering EPs transmit analog signals with amplitudes proportional to the query answers, allowing the noisy channel to act as a stabilizing mechanism. Due to the independence of the noise realizations, each transmission experience a new randomness, aligning with the widely adopted solution for controlling bias in the basic model of the ADA problem. This approach has also been applied in some recent FL studies to preserve data privacy for the EPs.

\section{Point-to-Point ADA without any communication constraint}
\label{sec:P2P_basic_ADA}
In this section, we present a key theorem from \cite{rogers2020guaranteed} regarding the basic model of ADA. The theorem establishes a relationship between the variance of Gaussian noise and the accuracy parameters of the answering mechanism. While our representation differs slightly from the original, it remains mathematically equivalent. In the subsequent sections, we apply this theorem to our proposed answering scheme for the federated adaptive data analysis.

\begin{thm}[Theorem 2.1. of \cite{rogers2020guaranteed}] \label{ThmRogers}
	Fix a Gaussian mechanism parameter  $\sigma>0$ and a desired confidence parameter $0<\beta<1$. The Gaussian mechanism can be used  to answer $ k $ statistical queries while satisfying $(\alpha,\beta)$-distributional accuracy, where $ \alpha $ is derived from the following expression:
	\begin{align}
		\label{eq_Rogers}
		\alpha=\max\lrbrc{ \sqrt{ \frac{2}{n\beta} . \min_{\lambda\in\left[ 0,1\right) } f\left(\lambda\right)} \,,\, \sqrt{8\sigma^2\ln{\frac{4k}{\beta}}}},
	\end{align}
	where
	\begin{align}
		f\left(\lambda\right)\triangleq\frac{1}{\lambda} \left(\frac{k}{n\sigma^2}-\ln\left(1-\lambda\right)\right) .
	\end{align}
\end{thm}
The first and second terms in \eqref{eq_Rogers} illustrate a tradeoff in $ \sigma $. The first term, which decreases with $\sigma$, is due to over-leakage regime, corresponds to the over-leakage regime, where a higher variance for noise is required to achieve a better level of differential privacy. The second term, which increases with $ \sigma $, represents the under-leakage regime, where a high variance of noise deteriorates data utility. In the following,  we derive a more manageable form for  \eqref{eq_Rogers}. First, we express the direct dependence of $ \alpha $ on  $ k $, $ n $ and $ \sigma^2 $. Then, we manipulate this function to express $ k $ as a function of $ n $ and $ \sigma^2$ for a given pair of $ (\alpha, \beta) $. This latter form, will be directly applicable in our problem setup in the following sections. 

\begin{figure}
	\centering
	\includegraphics[width=0.95\columnwidth,keepaspectratio]{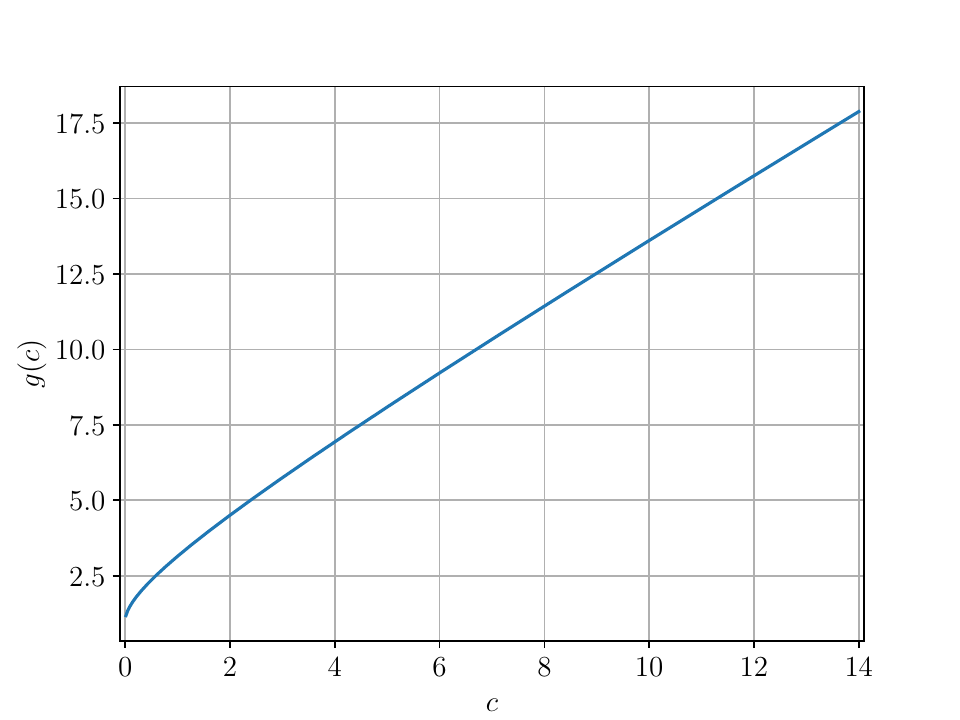}
	\caption{$ g(c) $ versus $ c $.}
	\label{fig_g_vs_c}
\end{figure}

\begin{lem}
	Consider the optimization problem $ \min_{\left[0,1\right)}{f(\lambda)} $ which is a part of the left expression in \eqref{eq_Rogers}. The solution of  this problem can be expressed in a closed form  as follows:
	\begin{align} 
		\label{g_c}
		\frac{W\left(-e^{-(c+1)}\right)\cdot\left(c + \ln\left(-W\left(-e^{-(c+1)}\right)\right) \right)}{1 + W\left(-e^{-(c+1)}\right)},
	\end{align}
	where $ W\left(\cdot\right)$ is the Lambert function and $c = \frac{k}{n\sigma^2}$. We will denote \eqref{g_c} by $ g(c) $.
\end{lem}
\begin{proof}
	A simple inspection shows that the minimum is not occurred in the extreme points. Hence,  to find the minimum, we take the derivative of \( f(\lambda) \) with respect to \( \lambda \) and set it to zero:
	\[
	f'(\lambda) = -\frac{1}{\lambda^2} \left( \frac{k}{n\sigma^2} - \ln(1-\lambda) \right) + \frac{1}{\lambda} \cdot \frac{1}{1-\lambda}=0.
	\]
	Consequently, this results in
	\[
	\frac{1}{1-\lambda} = \frac{1}{\lambda} \left( \frac{k}{n\sigma^2} - \ln(1-\lambda) \right).
	\]
	By rearranging the equation and defining $ c \triangleq  \frac{k}{n\sigma^2}$, we have:
	\[
	\frac{\lambda}{1-\lambda} + \ln(1-\lambda) = c.
	\]
	Defining  $ u \triangleq \frac{1}{1-\lambda} $ and consequently $ \lambda = 1 - \frac{1}{u} $, we obtain:
	\[
	\quad u - \ln(u) = c + 1.
	\]
	Setting $ v \triangleq -u $ and exponentiating both sides of the resulting equation:
	\[
	\quad v e^{v} = -e^{-(c+1)}.
	\]
	The last equation shows that $ v $ equals Lambert $W$ function evaluated at the point $ -e^{-(c+1)} $. The Lambert $W$ function is defined as the inverse of $ y = x e^x $ function, meaning it satisfies $ W(z)e^{W(z)} = z $. Finally, by substituting the equivalent values and performing a few rearrangements, we obtain:
	\begin{align} \label{LamStar}
		\lambda^{*} = 1 + \frac{1}{ W\left(-e^{-(\frac{k}{n\sigma^2}+1)}\right)}.
	\end{align}
	For negative real numbers in $ [-1/e,0) $, Lambert function has two branches, denoted by $ W_0 $ and $ W_{-1} $. Therefore, we need to determine which terms should appear in \eqref{LamStar}. The branch $ W_0 $ gives values between $ -1 $ and $ 0 $, while $ W_{-1} $ gives values in $ (-\infty,-1] $. On the other hand, due to the problem constraint $ 0 \leq \lambda <1 $, only the branch of $ W_{-1} $ is valid in \eqref{LamStar}. Furthermore, note that the variation domain of the argument of $ W $ in \eqref{LamStar} lies exactly in $ [-1/e,0) $, which is the domain of $ W_{-1} $. Finally, substituting $ \lambda^{*} $ in $ f(\lambda) $, gives \eqref{g_c}.
\end{proof}
Despite its complicated formulation in \eqref{g_c},  $g(c)$ behaves benignly with respect to $c$. Figure \ref{fig_g_vs_c} shows that for $ c>10 $, its growth closely follows a linear trend.

\begin{corol}
	Let $ \alpha $, $ \beta $, and $ n $ to be given. Then, $ k $ is derived from $ \sigma $ as follows:
	\begin{equation}\label{k_vs_sig}
		k(\sigma ; n, \alpha, \beta) = \min\lrbrc{ k_1 , k_2 },
	\end{equation}
	where $ k_1 $ and $ k_2 $ are defined as:
	\begin{align}\label{k_1_and_k_2}
		k_1(\sigma ; n, \alpha, \beta) &\triangleq n \sigma^2  g^{-1}\lrpar{\frac{n \alpha^2 \beta}{2}}, \\
		k_2(\sigma ; \alpha, \beta) &\triangleq \frac{\beta}{4} e^{\frac{\alpha^2}{8\sigma^2}}.
	\end{align}
\end{corol}
\begin{proof}
	Letting $ \sigma $ vary from $ 0 $ to larger values, we observe in \eqref{eq_Rogers} that, before reaching a critical value of $ \sigma $, the first term dominates. After this critical value, $ \alpha $ is determined by the second term. When $ \alpha $ is fixed, the functional dependence of  $ k $ on $ \sigma $ can be derived from either of these two terms. For each $\sigma$, the term that yields the smaller value of  $ k $ is the correct one.
\end{proof}
In the following sections, we set $ \alpha = 0.1 $ and $ \beta = 0.05 $, typical values in the literature. For these values, Figure \ref{fig_g_vs_c} shows that when $ n > 4 \times 10^4 $, the first term closely approximates
\begin{equation}\label{k_hat}
	\hat{k_1}(\sigma ; n, \alpha, \beta) = w \cdot \frac{n^2 \sigma^2 \alpha^2 \beta}{2} + b n\sigma^2,
\end{equation}
where $ w $ and $ b $ are some constants.

\section{Point-to-Point ADA in the Presence of a Gaussian Channel} \label{sec:P2P_ADA}
To clarify our idea, we begin with the case where there is only one EP answering the PS's adaptive queries via a Gaussian channel.  Specifically, at the
$i$'th question-answering round of this interactive procedure, the PS sends $q_i(\cdot)$ through an error-free channel and receives $a_i$ as the answer from the EP through a Gaussian channel with noise variance $\sigma^2$. As noted in \ref{ADA}, we aim to exploit the intrinsic channel noise as the randomization noise samples in the Gaussian answering mechanism. In this scheme, at each round, such as round $i$, the EP computes the empirical average of the query as
\begin{equation}
	q_i(S) = \frac{1}{n}\sum_{j=1}^{n}{q_i(x_j)},
\end{equation}
and then multiplies it by $A_{\mathsf{t}}$. The resulting signal is broadcast without additional coding, using analog Pulse Amplitude Modulation (PAM) with a maximum amplitude of $A_{\mathsf{t}}$. Using a matched filter, the PS detects a noisy version of the transmitted signal. Without applying any hard decision, it interprets the detected signal as $a_i$, the answer to its query. Therefore, we have:
\begin{equation}
	a_i = A_{\mathsf{t}}q_i(S) + z_i, 
\end{equation}
where $z_i \sim \Nc(0,\sigma^2)$ is the sampled Gaussian channel noise at the receiver. Assuming that the channel condition remains unchanged during the whole ADA procedure, it is clear that $n_i$'s are independent and identically distributed. Thus, the proposed scheme completely resembles the Gaussian answering mechanism in ADA, allowing us to apply Theorem \ref{ThmRogers} to this scenario. We use this theorem to adjust the tunable communication parameters such that the desired accuracy level is achieved. We assume that $(\alpha,\beta)$ has been given, and the dataset size, $n$, is fixed. Based on these parameters and as a function of $A_{\mathsf{t}}$, we can derive the maximum number of allowable queries for which \eqref{accdef} remains valid.

In the basic ADA model, the maximum value of the queries is assumed to be fixed at 1, and the variance of the artificial randomization noise is adjusted to achieve optimal performance. In contrast, in our proposed model, the noise variance is dictated by the physical layer conditions and is beyond our control. Instead, we tune the maximum amplitude of the query, denoted as $A_{\mathsf{t}}$, to achieve the best performance. Also, note that $\alpha$ in \eqref{accdef} is considered under the assumption that $|q_i(\cdot)| \leq 1$. Nonetheless, we continue using \eqref{accdef} by normalizing the maximum values of the answers to 1 when investigating this relation. More specifically, the validity condition of \eqref{accdef} for the situation $(\alpha = \alpha_0 \cdot A_0, \beta = \beta_0, A_{\mathsf{t}} = A_0, \sigma = \sigma_0)$ is entirely equivalent to the situation $(\alpha = \alpha_0, \beta = \beta_0, A_{\mathsf{t}} = 1, \sigma = \sigma_0 / A_0)$. Thus, by tuning $A_{\mathsf{t}}$, we implicitly calibrate the noise variance in the equivalent normalized model.

\begin{figure}
	\centering
	\includegraphics[width=0.95\columnwidth,keepaspectratio]{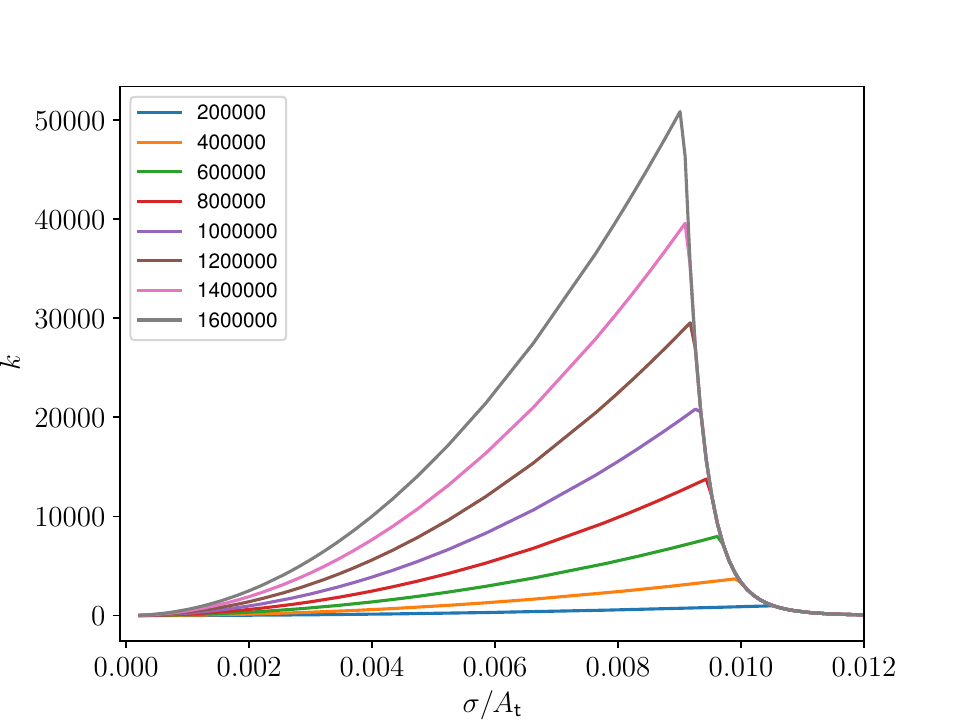}
	\caption{Achievable number of accurately answerable queries versus $\sigma/A_{\mathsf{t}}$ for $\alpha=0.1$ and $\beta=0.05$ and different values of $n$.}
	\label{fig_P2P_k_vs_PT}
\end{figure}

Therefore, to apply the theorem \ref{ThmRogers}, it's sufficient to replace $\sigma$ in  \eqref{k_vs_sig}, with $\frac{\sigma}{A_{\mathsf{t}}}$. In the figure \ref{fig_P2P_k_vs_PT} , $k$ is plotted in $\frac{\sigma}{A_{\mathsf{t}}}$, for some different values of $n$ and given desired accuracy parameters $\alpha = 0.1$ and $\beta=0.05$. 

\begin{figure}
	\centering
	\includegraphics[width=0.95\columnwidth,keepaspectratio]{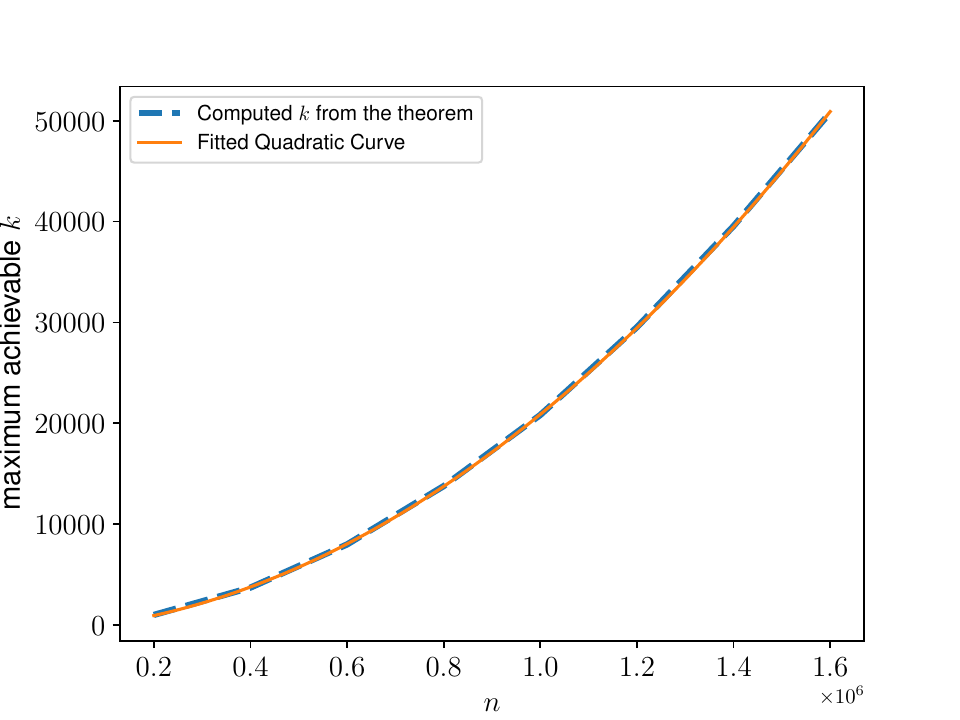}
	\caption{Maximum achievable number of accurately answerable queries versus $n$ for $\alpha=0.1$ and $\beta=0.05$.}
	\label{fig_Kmax_vs_n}
\end{figure}

Additionally, in Figure \ref{fig_Kmax_vs_n}, the maximum achievable number of queries is plotted against $n$. For each value of $n$, the optimal $ \sigma / A_{\mathsf{t}} $ that maximizes $k$ is selected. As shown in the figure, $k_{\text{max}}$ exhibits quadratic growth with respect to $n$. This observation aligns with the well-known result in the ADA literature that, in the worst-case setup, up to $\mathcal{O}(n^2)$ queries can be answered $(\alpha, \beta)$-accurately \cite{dagan2022bounded}.

Furthermore, Figure \ref{fig_P2P_k_vs_PT} shows that as $n$ increases, the optimum value of  $ \sigma/A_{\mathsf{t}} $ very slowly shifts to the left. This optimal point corresponds to the intersection of the quadratically increasing left curve and the exponentially decreasing right curve, as illustrated by $k_1$ and $k_2$ in \eqref{k_1_and_k_2}. For the dataset sizes considered in this study, the optimal value of $\sigma / A_{\mathsf{t}}$ approximately falls within the range $[0.008, 0.01]$. Furthermore, as $\sigma / A_{\mathsf{t}}$ approaches zero, the number of accurate queries also declines to near-zero levels. In this low-noise regime, the level of randomization becomes insufficient to prevent data leakage. As a result, the analyst can overfit to the dataset and violate the accuracy criterion in \eqref{accdef} with only a few queries. Thus, unlike standard communication scenarios, increasing transmission power in this setup can lead to inefficiency. On the other hand,  computation shows that in order to have a non-zero number of $(\alpha,\beta)$-accurate query-answers by this scheme, $\sigma/A_{\mathsf{t}}$ should be lower than 0.017 or equivalently $A_{\mathsf{t}}$ should be approximately 60 times greater than the standard deviation of the noise. This ratio, in the optimum situation, lies within the range of $[100, 125]$. Equivalently, a minimum SNR of 32 dB is required, with the optimum SNR falling between 37 and 38.9 dB. However, in the high-noise regime, given practical limitations on transmission power, improving $k$ may not be straightforward. In the next section, we demonstrate how a federation of edge answering points can address this issue."

\section{Distributed ADA in the Presence of a Gaussian Channel}
\label{sec:distributed_ADA}
In the previous section, we examined the point-to-point scenario to clarify the concept of effectively exploiting the channel noise. We showed that calibrating the amplitude of the transmitted signal can optimize the achievable number of accurately answered queries. Furthermore, we see that optimum $A_{\mathsf{t}}$ depends on the dataset size, $n$. 

On the other hand, although we have theoretically succeeded in utilizing noise as part of the adaptive query-answering procedure, the required SNR range may not always be feasible in certain communication scenarios. This challenge is particularly significant when the answerer is a personal device or a low-power sensor located at the network's edge and operating at a considerable distance from the receiver. As previously mentioned, this represents a plausible use case for our proposed setup. Additionally, we observed that for a fixed $A_{\mathsf{t}}$, the required dataset size $n$ grows at a rate of $\mathcal{O}(\sqrt{k})$. Consequently, ensuring more accurate (or private) rounds of data analysis would require the EP to answer queries based on increasingly larger datasets. However, it may be impractical for the EP to gather or store such large datasets. In this section, we demonstrate that both power and data limitations can be addressed within a distributed framework.

\begin{figure}
	\centering
	\includegraphics[width=0.95\columnwidth,keepaspectratio]{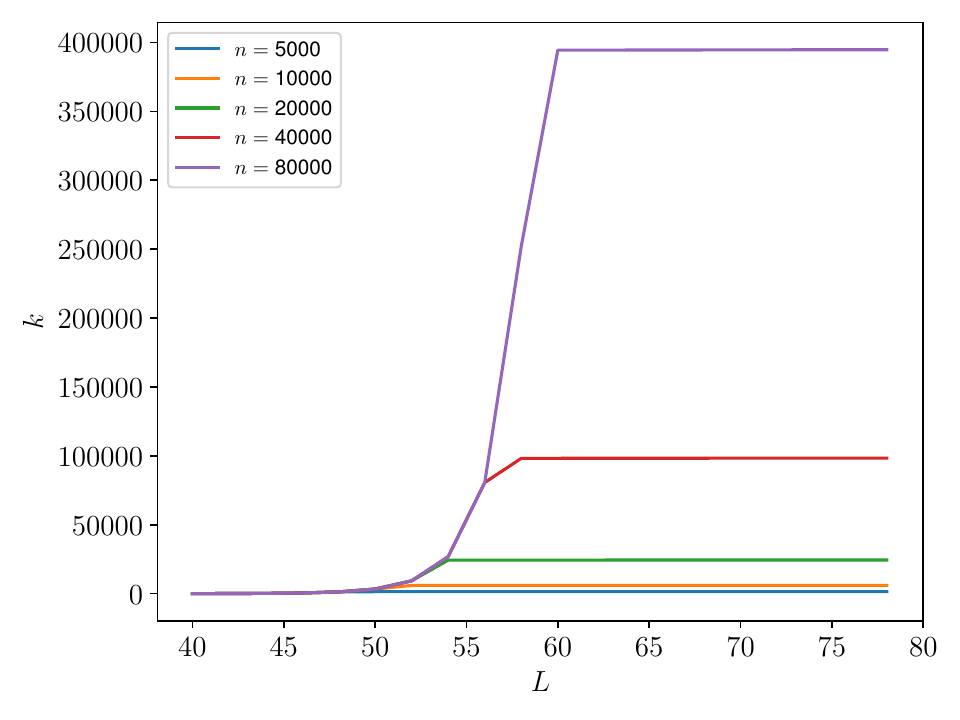}
	\caption{{Maximum number of accurately answered queries versus $\sigma/A_{\mathsf{t}}$} for $\alpha=0.1$ and $\beta=0.05$ and different values of $n$.}
	\label{fig_dist_gauss_k_vs_L}
\end{figure}

We consider a situation where $L$ EPs simultaneously transmit their answers to the CP's adaptive queries through an analog Gaussian multiple access channel (MAC). For simplicity, we assume that the dataset sizes of all EPs are identical, i.e., $n_l = n_0$ for all $l \in {1, 2, \dots, L}$. Similar to the previous section, each EP, like $\Mc_l$, computes $q_{i}(S_l) = \frac{1}{n_0}\sum_{j=1}^{n_0}{q_i(x_{j,l})}$ and transmit $ A_{\mathsf{t}} q_{i}(S_l) $ without any coding.  Because of the additive nature of the air interface, the receiver gets a noisy version of the aggregated signals. So, the received signal at the CP will be $a_i = A_{\mathsf{t}}\sum_{l=1}^{L}{q_i(S_l)} + z_i$, where $z_i \sim \Nc(0,\sigma^2)$. To establish equivalence with the point-to-point case, we consider $a_i / L$ in the following. Substituting $q_i(S_l)$ yields:
\begin{equation}
	\frac{a_i}{L} = \frac{A_{\mathsf{t}}}{L n_0}\sum_{l=1}^{L}{\sum_{j=1}^{n_0}{q_i(x_{j,l})}} + \frac{z_i}{L}.	 
\end{equation}
The above equation completely resembles a point-to-point scenario with $n_{\text{eq}} = L n_0$, $\sigma_{\text{eq}}=\frac{\sigma}{L}$ and $A_{T,eq}=A_{\mathsf{t}}$. 
We observe that with this cooperative transmission scheme, both the equivalent dataset size and the noise variance are improved by a factor of $L$. Using these equivalent parameters, the results from the previous section can be applied to this scenario as well. Thus, we can use \eqref{k_vs_sig} with the corresponding $n_{\text{eq}}$ and $\sigma_{\text{eq}}$. In this case, we fix $\sigma/A_{\mathsf{t}} = 0.5$ and derive $k$ as a function of $L$ for different values of $n$. Based on the results from the previous section, we know that $L$ must be sufficiently large to ensure that $\sigma_{\text{eq}}$ is less than $0.017$, thereby yielding a positive value for $k$. Figure \ref{fig_dist_gauss_k_vs_L} illustrates the results for this scenario. The parameters $\alpha$ and $\beta$ remain the same as in the previous case.
We observe that for each $n$, before reaching a critical number of EPs, $k$ grows exponentially with $L$. This is due to the under-leakage regime and is related to $k_2$ in \eqref{k_1_and_k_2}. The critical value of $L$ shifts slowly to the right as $n$ increases. After this critical $L$, the growth of $k$ becomes very slow, and in Figure \ref{fig_dist_gauss_k_vs_L}, it appears nearly constant. This is because, beyond the critical $L$, $k$ follows the $k_1$ function from \eqref{k_1_and_k_2}. As mentioned earlier, $k_1$ can be approximated by $\hat{k_1}$ in \eqref{k_hat}. In this function, the quadratic term becomes strongly dominant for large $n$. In this quadratic term, $ n^2\sigma^2 $ is determining. On the other hand, in the distributed scenario, we havebb
\begin{align}
	n_{\text{eq}} \sigma _{\text{eq}} = (L n) (\sigma / L) = \sigma,
\end{align}
which does not depend on $ L $. This argument also provides a perspective for addressing this inefficient growth stoppage, which is discussed in the following subsection.

\subsection{Improving $ k $ in $ L $}
\begin{figure}
	\centering
	\includegraphics[width=0.95\columnwidth,keepaspectratio]{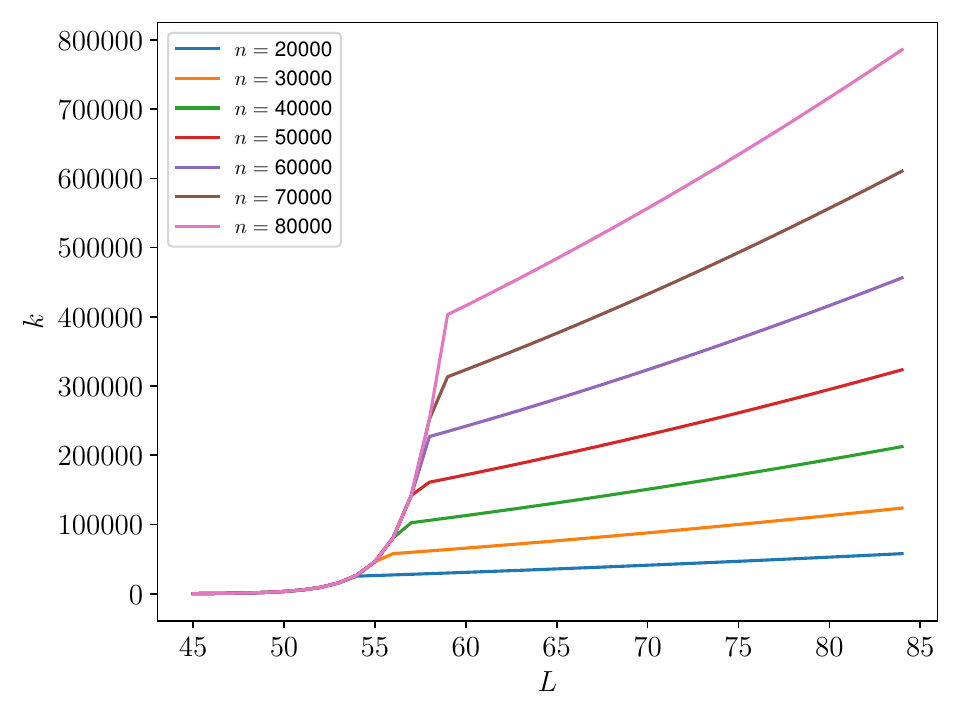}
	\caption{{Maximum number of accurately answered queries versus $\sigma/A_{\mathsf{t}}$} for $\alpha=0.1$ and $\beta=0.05$ and different values of $n$.}
	\label{fig_imp_dist_g_k_vs_L}
\end{figure}

Considering the descriptions in Figure \ref{fig_P2P_k_vs_PT} about the optimum value of $ \sigma/A_{\mathsf{t}} $, EPs in the distributed scenario need to decrease $ A_{\mathsf{t}} $ as $ L $ exceeds the critical value. For large $ n_{\text{eq}} $, the optimum value of $ \sigma_{\text{eq}}/A_{\mathsf{t}} $ is approximately $ 0.008 $. The exact value can be derived by setting $ k_1 = k_2 $ in \eqref{k_vs_sig}. Denoting this value as $ s_{\text{opt}}(n_{\text{eq}}) $, the optimum value of $ \sigma / A_{\mathsf{t}} $ will be $ L s_{\text{opt}}(n_{\text{eq}}) $. In other words, EPs should tune their $ A_{\mathsf{t}} $ to $ \frac{\sigma }{L s_{\text{opt}}(L n)} $. As noted in the previous section, $ s_{\text{opt}} $ changes very slowly. Therefore, we can approximately state that in the optimum regime, $ A_{\mathsf{t}} \propto 1/L $. Figure \ref{fig_imp_dist_g_k_vs_L} illustrates the resulting improvement in the growth of $ k $ after optimizing $ A_{\mathsf{t}} $.

\section{Future Work and Conclusion}
\label{sec:conclusion}
In this work, after refining a previous result in ADA to better suit our analyses, we showed that Gaussian channel noise can be leveraged as an opportunity when the basic model of ADA is considered. Of course, in this situation, the minimum required SNR was significant. Then, we demonstrated that the federation of edge answering points can linearly improve the equivalent noise variance and total dataset size when the shared medium is a Gaussian MAC.

One important extension for further study could involve considering the fading channel, which represents a more realistic model. In this case, each EP's transmission undergoes a multiplicative distortion before being received by the CP. In other words, the received signal during the $i$-th answering round would be as follows:
\begin{equation}
	Y_i = \sum_{l=1}^{L}{h_{il}X_{il}+z_i},
\end{equation}
in which $ X_{il} $ represents the signal transmitted by the $l$-th EP, and $ h_{il} $ denotes the fading coefficient of the wireless channel between this EP and the CP. Based on the availability of CSI, various scenarios can be analyzed: CSI known only at the transmitter, CSI known only at the receiver, and CSI unknown at both ends. A primary challenge in the fading scenario is that the unweighted aggregation of transmissions is no longer inherently feasible.

\bibliographystyle{IEEEtran}
\bibliography{Reference}

\begin{thebibliography}{10}
\providecommand{\url}[1]{#1}
\csname url@samestyle\endcsname
\providecommand{\newblock}{\relax}
\providecommand{\bibinfo}[2]{#2}
\providecommand{\BIBentrySTDinterwordspacing}{\spaceskip=0pt\relax}
\providecommand{\BIBentryALTinterwordstretchfactor}{4}
\providecommand{\BIBentryALTinterwordspacing}{\spaceskip=\fontdimen2\font plus
\BIBentryALTinterwordstretchfactor\fontdimen3\font minus
  \fontdimen4\font\relax}
\providecommand{\BIBforeignlanguage}[2]{{%
\expandafter\ifx\csname l@#1\endcsname\relax
\typeout{** WARNING: IEEEtran.bst: No hyphenation pattern has been}%
\typeout{** loaded for the language `#1'. Using the pattern for}%
\typeout{** the default language instead.}%
\else
\language=\csname l@#1\endcsname
\fi
#2}}
\providecommand{\BIBdecl}{\relax}
\BIBdecl

\bibitem{dwork2015preserving}
C.~Dwork, V.~Feldman, M.~Hardt, T.~Pitassi, O.~Reingold, and A.~L. Roth,
  ``Preserving statistical validity in adaptive data analysis,'' in
  \emph{Proceedings of the forty-seventh annual ACM symposium on Theory of
  computing}.\hskip 1em plus 0.5em minus 0.4em\relax ACM, 2015, pp. 117--126.

\bibitem{dwork2015generalization}
C.~Dwork, V.~Feldman, M.~Hardt, T.~Pitassi, O.~Reingold, and A.~Roth,
  ``Generalization in adaptive data analysis and holdout reuse,'' in
  \emph{Advances in Neural Information Processing Systems}, 2015, pp.
  2350--2358.

\bibitem{rogers2020guaranteed}
R.~Rogers, A.~Roth, A.~Smith, N.~Srebro, O.~D. Thakkar, and B.~Woodworth,
  ``Guaranteed validity for empirical approaches to adaptive data analysis,''
  in \emph{International Conference on Artificial Intelligence and Statistics},
  2020, pp. 2830--2840.

\bibitem{dwork2014algorithmic}
C.~Dwork, A.~Roth \emph{et~al.}, ``The algorithmic foundations of differential
  privacy,'' \emph{Foundations and Trends{\textregistered} in Theoretical
  Computer Science}, vol.~9, no. 3--4, pp. 211--407, 2014.

\bibitem{yang2021revisiting}
H.~H. Yang, Z.~Chen, T.~Q. Quek, and H.~V. Poor, ``Revisiting analog
  over-the-air machine learning: The blessing and curse of interference,''
  \emph{IEEE Journal of Selected Topics in Signal Processing}, 2021.

\bibitem{amiri2020federated}
M.~M. Amiri and D.~G{\"u}nd{\"u}z, ``Federated learning over wireless fading
  channels,'' \emph{IEEE Transactions on Wireless Communications}, vol.~19,
  no.~5, pp. 3546--3557, 2020.

\bibitem{mohamed2021privacy}
M.~S.~E. Mohamed, W.-T. Chang, and R.~Tandon, ``Privacy amplification for
  federated learning via user sampling and wireless aggregation,'' \emph{IEEE
  Journal on Selected Areas in Communications}, vol.~39, no.~12, pp.
  3821--3835, 2021.

\bibitem{sonee2021wireless}
A.~Sonee, S.~Rini, and Y.-C. Huang, ``Wireless federated learning with limited
  communication and differential privacy,'' \emph{arXiv preprint
  arXiv:2106.00564}, 2021.

\bibitem{seif2020wireless}
M.~Seif, R.~Tandon, and M.~Li, ``Wireless federated learning with local
  differential privacy,'' in \emph{2020 IEEE International Symposium on
  Information Theory (ISIT)}.\hskip 1em plus 0.5em minus 0.4em\relax IEEE,
  2020, pp. 2604--2609.

\bibitem{xiaowen2021optimized}
C.~Xiaowen, Z.~Guangxu, X.~Jie, W.~Zhiqin, and C.~Shuguang, ``Optimized power
  control design for over-the-air federated edge learning,'' \emph{arXiv
  preprint arXiv:2106.09316}, 2021.

\bibitem{koda2020differentially}
Y.~Koda, K.~Yamamoto, T.~Nishio, and M.~Morikura, ``Differentially private
  aircomp federated learning with power adaptation harnessing receiver noise,''
  in \emph{GLOBECOM 2020-2020 IEEE Global Communications Conference}.\hskip 1em
  plus 0.5em minus 0.4em\relax IEEE, 2020, pp. 1--6.

\bibitem{koda2021airmixml}
Y.~Koda, J.~Park, M.~Bennis, P.~Vepakomma, and R.~Raskar, ``Airmixml:
  Over-the-air data mixup for inherently privacy-preserving edge machine
  learning,'' \emph{arXiv preprint arXiv:2105.00395}, 2021.

\bibitem{liu2020privacy}
D.~Liu and O.~Simeone, ``Privacy for free: Wireless federated learning via
  uncoded transmission with adaptive power control,'' \emph{IEEE Journal on
  Selected Areas in Communications}, vol.~39, no.~1, pp. 170--185, 2020.

\bibitem{zhang2021turning}
Z.~Zhang, G.~Zhu, R.~Wang, V.~K. Lau, and K.~Huang, ``Turning channel noise
  into an accelerator for over-the-air principal component analysis,''
  \emph{arXiv preprint arXiv:2104.10095}, 2021.

\bibitem{OBDAgunduz}
G.~Zhu, Y.~Du, D.~Gündüz, and K.~Huang, ``One-bit over-the-air aggregation
  for communication-efficient federated edge learning: Design and convergence
  analysis,'' \emph{IEEE Transactions on Wireless Communications}, vol.~20,
  no.~3, pp. 2120--2135, 2021.

\bibitem{shao2021federated}
Y.~Shao, D.~G{\"u}nd{\"u}z, and S.~C. Liew, ``Federated edge learning with
  misaligned over-the-air computation,'' \emph{IEEE Transactions on Wireless
  Communications}, 2021.

\bibitem{amiri2020blind}
M.~M. Amiri, T.~M. Duman, D.~Gunduz, S.~R. Kulkarni, and H.~V. Poor, ``Blind
  federated edge learning,'' \emph{arXiv preprint arXiv:2010.10030}, 2020.

\bibitem{amiri2020machine}
M.~M. Amiri and D.~G{\"u}nd{\"u}z, ``Machine learning at the wireless edge:
  Distributed stochastic gradient descent over-the-air,'' \emph{IEEE
  Transactions on Signal Processing}, vol.~68, pp. 2155--2169, 2020.

\bibitem{lee2021over}
C.-Z. Lee, L.~P. Barnes, and A.~{\"O}zg{\"u}r, ``Over-the-air statistical
  estimation,'' \emph{IEEE Journal on Selected Areas in Communications}, 2021.

\bibitem{dagan2022bounded}
Y.~Dagan and G.~Kur, ``A bounded-noise mechanism for differential privacy,'' in
  \emph{Conference on Learning Theory}.\hskip 1em plus 0.5em minus 0.4em\relax
  PMLR, 2022, pp. 625--661.

\end{thebibliography}

\end{document}